\documentclass[runningheads]{llncs}
\usepackage[T1]{fontenc}
\usepackage{graphicx}
\usepackage{algpseudocode}
\usepackage{algorithm}
\usepackage{amsmath}
\usepackage{amssymb}
\usepackage{caption}
\usepackage{subcaption}
\usepackage{xcolor}

\algdef{SE}[REPEATN]{RepeatN}{End}[1]{\algorithmicrepeat\ #1 \textbf{times}}{\algorithmicend}

\begin{document}
%
\title{Multi-dimensional Parameter Space Exploration for Streamline-specific Tractography}
\titlerunning{Parameter Space Exploration for Streamline-specific Tractography}
%
\author{Ruben Vink\inst{1} \and
Anna Vilanova\inst{1} \and Maxime Chamberland\inst{1}}
%
\authorrunning{R. Vink et al.}
%
\institute{Department of Computer Science and Mathematics, Eindhoven University of Technology, Eindhoven, The Netherlands}
\maketitle              
\begin{abstract}
One of the unspoken challenges of tractography is choosing the right parameters for a given dataset or bundle. In order to tackle this challenge, we explore the multi-dimensional parameter space of tractography using streamline-specific parameters (SSP). We 1) validate a state-of-the-art probabilistic tracking method using per-streamline parameters on synthetic data, and 2) show how we can gain insights into the parameter space by focusing on streamline acceptance using real-world data. We demonstrate the potential added value of SSP to the current state of tractography by showing how SSP can be used to reveal patterns in the parameter space.

\keywords{Probabilistic Tractography \and Streamline-specific parameters \and Parameter Space Exploration.}
\end{abstract}
%
%
\section{Introduction}
Diffusion MRI is a non-invasive, in vivo imaging technique that can probe the underlying white matter architecture using tractography. Most tractography algorithms \cite{FACT,tensor_det,tensor_prob,iFOD2} can be classified as either deterministic or probabilistic \cite{Descoteaux09}. In both cases, many parameters (such as step size, angle, or tracking thresholds) have to be manually set (and fixed across subjects and bundles) beforehand. However, due to anatomical heterogeneity, it is sometimes necessary to adjust these parameters at the individual level, especially in pathological cases \cite{ensemble-tractography,maxi-2014}. Machine learning approaches have been proposed in healthy brains \cite{TOM-tracking} and more recently for diseased brains \cite{attractive}. The commonality of existing methods puts a focus on defining bundle-specific parameters (BSP) to improve tracking results \cite{ETZ-pipeline,BST}. The main challenge arises when the \textit{generally accepted} parameters start to fail. A potential solution would be to look at streamline-specific parameters (SSP). Here, we explore the viability of SSP in existing tractography algorithms with the main goal of improving tracking results. We use an in-house tracking algorithm based on general probabilistic tractography algorithms that allows the per-streamline parameter sampling and the extraction of the information required for our analysis. 

Ensemble tractography \cite{ensemble-tractography,ensemble-tractography-pipeline} is a similar method that was proposed to solve the same issues. However, with ensemble tractography $N$ sets of parameters are used to generate $N$ tractograms -- each with many streamlines -- which are merged at a later stage and an optimised set of parameters is obtained. SSP takes $M$ samples of a given parameter space and generates $M$ streamlines, resulting in a much denser sampling of the parameter space. Using SSP to obtain insights (e.g. optimised parameter settings) proves to be a complex task due to the vast number of parameters and the extensive resulting parameter space. Parameters can be categorized as numerical (e.g. step size, maximum angle between successive steps, number of samples) or spatial (e.g. seed regions, tracking masks, inclusion regions). Aside from these parameters, there are also \textit{rule-based} parameters that are used to decide whether to accept or reject a generated streamline (e.g., filtering parameters). 
In practice, each algorithm has its own additional set of parameters, and that is without taking into account parameters related to anatomical priors \cite{ACT,reducing-biases,BST,MAGNET} .

In this study, we present an analysis of a subset of the multi-dimensional parameter space to explore the viability of future automatic exploration methods. An in-house implementation of a tracking algorithm is validated and used for the analysis.

\section{Methods}

\subsection{Tracking Algorithm}
Streamline specific parameter tracking (SSPT) is a probabilistic tracking method similar to iFOD2 as implemented in MRtrix3 \cite{iFOD2,MRtrix3} and is intended to serve as a representation of probabilistic tracking algorithms, relying only on the Fiber Orientation Distribution functions (FODs)\cite{CSD} and input ROIs. 
At a given position, SSPT picks a random direction $\mathbf{d'}$ in a cone around the current tracking direction $\mathbf{d}$. 
Much like iFOD2, SSPT then calculates an approximation of the joint probability of a step in direction $\mathbf{d'}$ by extrapolating a path in that direction and evaluating the FODs along the path using a discretized sphere and precomputed spherical harmonics.
SSPT does this by simply sampling 4 points linearly between $\mathbf{p}$ and $\mathbf{p'}$, whereas iFOD2 places these points along a circular arc such that the current direction is tangent to the start of the arc, and the sampled direction is tangent to the end of the arc. 
SSPT then chooses a direction $\mathbf{d'}$ at random out of at most 4 potential directions\footnote{The exception to this is the very first direction, since this is sampled over the entire sphere. A fixed 32 samples are taken instead to increase the chances a valid starting direction is found when possible.} where directions that do not pass the FOD threshold are not considered.
Each direction is weighted by the approximation of the joint probability.
The choice was made to use this approach over rejection sampling like in iFOD2 because of the biases rejection sampling can add in the interpretation of the parameter distributions. 
For example, one set of parameters might need many rejections to find a specific trajectory, whereas another set of parameters could find the same trajectory with far fewer samples. 
Solving this in post-processing would slow down the exploration too much for practical use. 
Pseudocode outlining SSPT is given in Algorithm~\ref{alg:SSPT}. Furthermore, SSPT can make use of binary masks for seeding, an arbitrary number of inclusion zones (both `and' and `or'), and exclusion zones. Additionally, backtracking was also added as done in \cite{ACT,reducing-biases}. The number of times it tries to backtrack is capped at a constant value (see Table~\ref{tbl:fixed_all}), and works by simply taking a single step back (instead of forwards) on the currently tracked path.
\begin{algorithm}[t]
\caption{Streamline Specific Parameter Tracking}
\hspace*{\algorithmicindent} 
\textbf{Input:} Random $\mathtt{seed}$, inclusion/exclusion ROIs, FODs\\
\hspace*{\algorithmicindent} 
\textbf{Output:} Tracking information for 1 seed and random set of parameters.
\label{alg:SSPT}
\begin{algorithmic}[1]
\State Generate random set of parameters
\State $\mathtt{points} \gets \{\mathtt{seed}\}$
\State Let $\mathbf{d}$ be a valid start direction if possible, otherwise terminate
\While{$|\mathtt{points}| \cdot \mathtt{step\_size} < \mathtt{max\_length}$}
    \State $\mathtt{candidates} \gets \emptyset$
    \State $\mathtt{weights} \gets \emptyset$
    \RepeatN{$\mathtt{n\_samples}$}\Comment{From Table~\ref{tbl:fixed_all}: $\mathtt{n\_samples} = 4$}
        \State Sample random direction $\mathbf{d'}$ in cone around $\mathbf{d}$.
        \State Let $\mathbf{x_i}$ be points along straight path from $\mathbf{p}$ to $\mathbf{p'} := \mathbf{p} + \mathtt{step\_size}\cdot\mathbf{d'}$
        \If{$\forall x_i$, $\mathtt{eval\_fod}(\mathbf{x_i}, \mathbf{d'}) > \mathtt{fod\_threshold}$}
            \State Add $\mathbf{d'}$ to $\mathtt{candidates}$ and add $\prod_{x_i} x_i$ to $\mathtt{weights}$.
        \EndIf{}
    \End
    
    \State
    \If{$|\mathtt{candidates}| > 0$} 
    \State Take step in new direction sampled from $\mathtt{candidates}$ weighted by $\mathtt{weights}$
    \Else
        \State Backtrack or terminate
    \EndIf{}
    \State
    \State Update status regarding inclusion regions
    \State
    \If{Current point in exclusion region}
        \State Backtrack or terminate
    \EndIf{}
\EndWhile{}
\State \Return tracked points, generated parameters, and other tracking information
\end{algorithmic}
\end{algorithm}

Finally and most importantly, SSPT allows for tracking each streamline with its own set of parameters sampled from user specified distributions. 
The output of a single iteration of the algorithm is a data structure containing information about the tracking such as which parameters were used, whether a streamline was found, how many backtracking attempts were performed, and how long it took to compute. 
Additionally, flags describing the reason for failing are included as well.
It is important to note that a streamline is not necessarily part of this data structure, since SSPT can complete without producing a streamline -- and does so more often than not by design. The information obtained in this way is used for the analysis of our experiments. The specific parameter distributions used for each experiment are discussed in their corresponding sections.

\subsubsection{Parameters}
The SSPT algorithm has a few parameters that are kept fixed for every experiment (see Table~\ref{tbl:fixed_all}). Those were pre-determined and deliberately kept the same to reduce the size of the parameter space. Additionally, the target number of streamlines to find was set per experiment. This study focuses on two varying parameters: the step size and the radius of curvature. The radius of curvature is sampled uniformly rather than the angle of the cone, since a larger step size requires a larger angle to produce the same curvature. The relation of radius of curvature $r$, step size $\Delta x$, and angle $\alpha$ is defined as follows:
\begin{equation}
    r = \frac{\Delta x}{\sin{\frac{\alpha}{2}}}.
\end{equation}
The cone angle is calculated from this relation after uniformly sampling between minimum and maximum radius.

\begin{table}[hbtp]
\caption{Fixed tracking parameters over all experiments}
\label{tbl:fixed_all}
\centering
    \begin{tabular}{|l|c|l|}
        \hline
        Parameter name & Value & Description \\
        \hline
        $\mathtt{SH\_resolution}$ & 4 & Number of subdivisions of discretized sphere. \\
        $\mathtt{backtrack\_lim}$ & 64 & Maximum number of times backtracking occurs.\\
        $\mathtt{intermediate\_steps}$ & 4 & Number of steps for computing joint probabilities. \\
        $\mathtt{n\_samples}$ & 4 & Number of directions that are always sampled. \\
        $\mathtt{seed\_samples}$ & 32 & Number of directions sampled at the seed point. \\
        $\mathtt{FOD\_threshold}$ & 0.1 & Minimum FOD value in a given direction for a valid step. \\
        \hline
    \end{tabular}
\end{table}

\subsubsection{Automatic Validation of Tracking Algorithm}
For the validation of the tracking algorithm, preprocessed data from the ISMRM2015 challenge's recent update \cite{main-ismrm-paper} was used. Bidirectional tracking on the reconstructed FODs was performed with seeding in a white matter mask with the grey matter white matter interface as inclusion region. Two million streamlines were generated using a step size between 0.4mm and 0.6mm and a radius of curvature between 0.75mm and 1.0mm. These ranges were selected after a single test run with larger ranges, where these parameter ranges showed a good balance between high acceptance rate and speed. An FOD amplitude threshold of 0.1 was used. The updated checker script provided by Renauld et al\cite{ismrm-2023-update} was then used to score the resulting tractogram for comparison purposes. Our method used the same masks as the `WM seeding + PFT tracking' entry and ran on all 24 cores of an Intel i7-13700K with 64GB of RAM.

\subsection{Streamline-specific Experiments}
The data used for the streamline-specific experiments consisted of one healthy subject from the minimally preprocessed Human Connectome Project (HCP) \cite{HCP}, and two clinical datasets provided by the Elisabeth TweeSteden Hospital (ETZ) in Tilburg, The Netherlands.
The clinical datasets used were preprocessed using by the tractography pipeline currently in use at ETZ \cite{ETZ-pipeline}. 
The FODs and masks generated by the pipeline are directly used with no further processing.

\textbf{Patient A} has a Glioblastoma Multiforme, WHO grade 4, in the vicinity of the angular gyrus near the arcuate fasciculus (AF).
\textbf{Patient B} has an oligodendroglioma, WHO grade 2, in the postcentral gyrus near the corticospinal tract (CST). Both datasets were acquired with a Philips Achieva 3T MRI scanner ($b = 1500$ s/mm$^2$, 50 diffusion-weighting directions, six b0 s/mm$^2$ images, 2 mm isotropic voxel size, TE/TR/echo spacing 87/8000/0.2 ms). The HCP dataset was acquired at a 1.25 mm isotropic voxel size, and only the b0 and $b = 3000$ s/mm$^2$ were used, in 90 directions.


\subsubsection{Per-streamline parameter distributions}

The following experiment shows how SSPT can provide detailed information during the tracking process of a specific bundle in cases where the fixed-parameter method presents shortcomings. This means that either 1) not enough streamlines can be found within the allowed number of seeds or time; or that 2) parts of the bundle that are expected to be tracked are not present in the output. Two specific bundles were selected to highlight the usefulness of SSPT: the corticospinal tract (CST), and the arcuate fasciculus (AF). The CST is used to show how, within a bundle, streamline specific parameters can be used to target specific parts of the bundle (e.g., fanning) in a healthy subject. To do so, the bundle is segmented into sub-parts using Quickbundles \cite{quickbundles}. The step size is sampled uniformly between 0.2 and 2 times the voxel size (i.e., 1.25 mm$^3$). Going larger than 2 times the voxel size can cause overshoot. The lower bound is in line with literature \cite{mrtrix2012}. The radius of curvature is sampled uniformly between 2 mm and 100 mm. Lower radii result in streamlines that are no longer anatomically viable, and any radii greater than 100mm all end up with many of the same -- almost straight -- streamlines. 

\subsubsection{Clinical datasets}
To assess whether the method can be applied in pathological cases, the AF in Patient A and the CST in patient B are reconstructed using SSPT. The results of SSPT are compared with iFOD2 \cite{iFOD2} (manual specification of bundle-specific parameters) and TractSeg using Tract Orientation Map tracking\cite{TOM-tracking} (automated approach). For iFOD2 the bundle-specific parameters as determined by Meesters et al. \cite{ETZ-pipeline} were used and for TractSeg the default parameters were used. In the end, we qualitatively evaluate how tracking is affected by the tumor region in all three approaches. Light filtering (<1\% of streamlines) based on fibre-to-bundle coherence \cite{FBC} was applied to the results of iFOD2 for visualization purposes only.
\section{Results}
\subsection{Validation of Tracking Algorithm}
The SSPT algorithm was validated using the ISMRM2015 challenge. The scores achieved were in line with previous scores and therefore SSPT was used as presented in the rest of the experiments. Table~\ref{tbl:ismrm-scores} shows the scores of our method compared to the results presented by Renauld et al\cite{ismrm-2023-update}. It is important to note that the goal is to show that SSPT can serve as a \textit{representative} for probabilistic tracking methods, and not necessarily as an \textit{alternative}. For this reason the exact same FODs and ROIs were used as Renauld et al. and parameters were not optimised to maximise score.

\begin{table}
    \caption{ISMRM2015 scores using the 2023 checker script by Renauld et al. \cite{ismrm-2023-update}}
    \label{tbl:ismrm-scores}
    \centering
    \begin{tabular}{|c|c|c|c|c|c|}
    \hline
    Order & VB (out of 21) & VS & mean OL & mean ORn & mean F1 \\
    \hline
    WM seeding + local tracking & 19 & \textbf{42.0}\% & \textbf{82.6}\% & 121.3\% & 51.7\% \\
    WM seeding + PFT tracking & 19 & 33.5\% & 68.7\% & 54.1\% & 58.2\% \\
    Interface seeding + PFT tracking & 19 & 28.8\% & 64.9\% & 44.5\% & \textbf{58.4}\% \\
    \hline
    SSPT (Ours) & \textbf{20} & 32.2\% & 55.8\% & \textbf{33.9}\% & 56.2\% \\
    \hline
    \end{tabular}
\end{table}
\subsection{Parameter Space Exploration}

\begin{figure}[htb!]
\centering
\includegraphics[width=\textwidth]{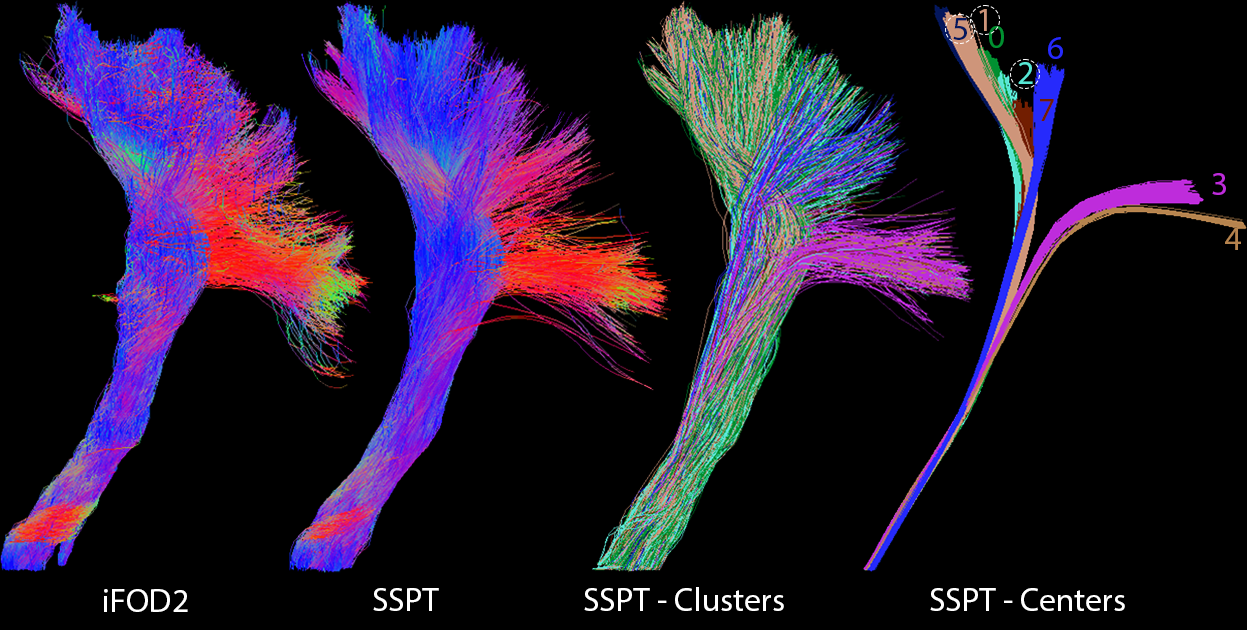}
\caption{Trackings of the left CST of the healthy HCP subject. From left to right; results from iFOD2 with the default parameters for the pipeline \cite{ETZ-pipeline}. Then the result of SSPT with the same ROIs. The last two images show a representation of the cluster representatives. The circled cluster numbers correspond to the data shown in Figure~\ref{fig:params_clusters}.}
\label{fig:cst}
\end{figure}
\begin{figure}[htb!]
\centering
\includegraphics[width=\textwidth]{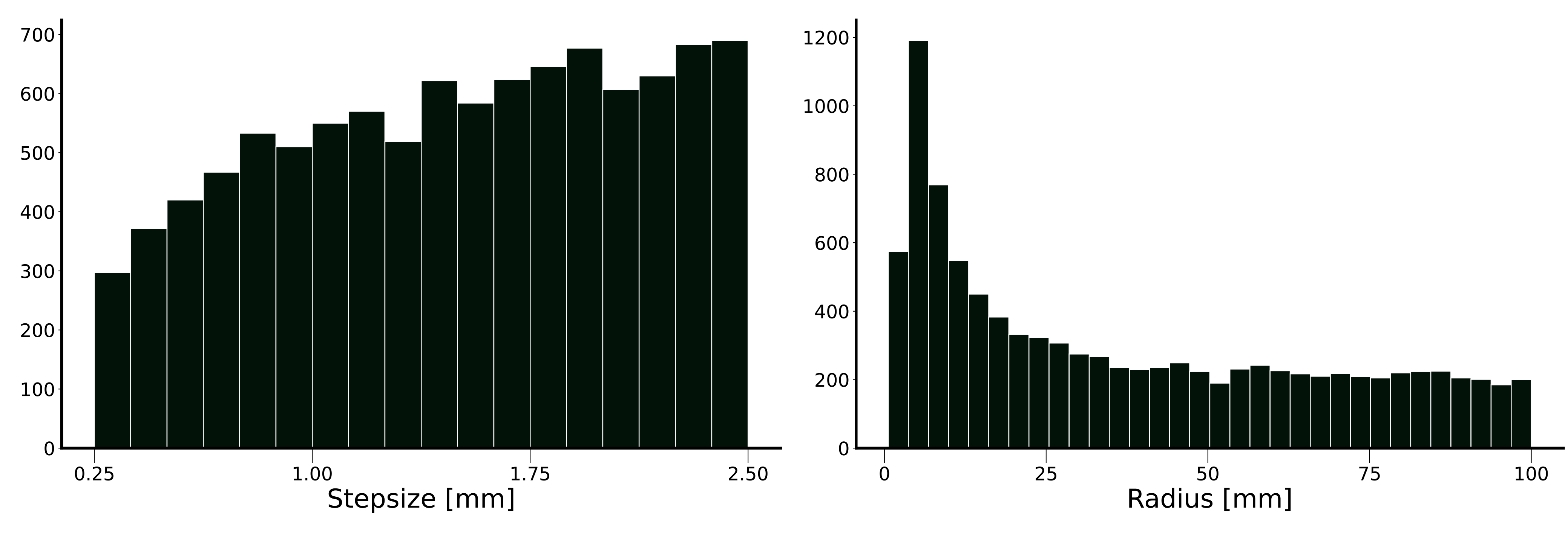}
\caption{Histograms of the amount of streamlines out of 10000 that were accepted with a specific parameter in the left CST of the healthy HCP subject.}
\label{fig:params_all}
\end{figure}
\begin{figure}[htb!]
\centering
\includegraphics[width=\textwidth]{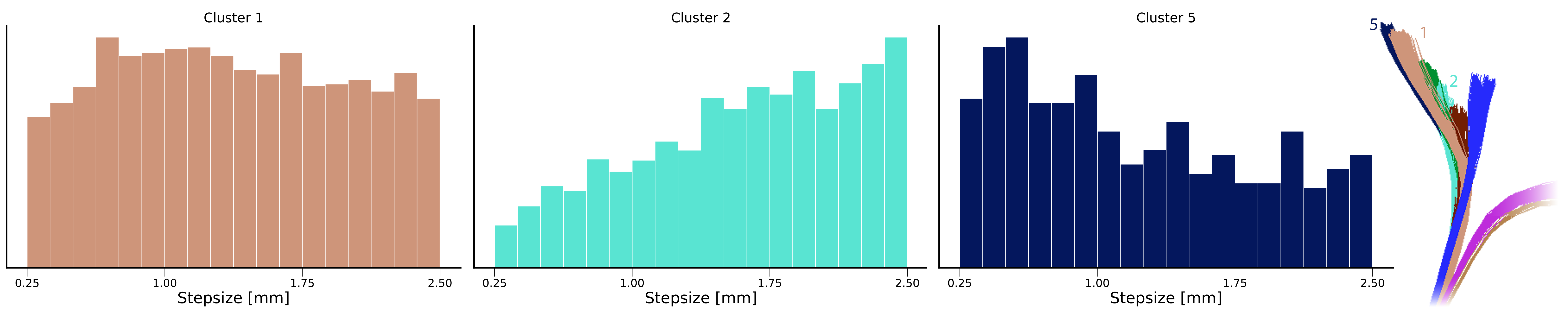}
\caption{Histograms for specific clusters as shown in Figure~\ref{fig:cst} (left CST of healthy HCP subject). On the right a close up of the cluster representatives is shown.}
\label{fig:params_clusters}
\end{figure}

The results of tracking the left CST in the healthy HCP dataset are shown in Figure~\ref{fig:cst}. Figure~\ref{fig:params_all} shows the distribution of parameters among all accepted streamlines, and Figure~\ref{fig:params_clusters} shows the distribution of parameters for three clusters.
By choosing parameters with high success rate (i.e., 3mm for the radius of curvature and 0.625mm for the step size), the running time is reduced from 3 minutes to 20 seconds. Additionally, Figure~\ref{fig:params_clusters} shows a clear difference between a cluster that does not seem to have a clear bias (cluster 1), and two clusters that have opposing biases (clusters 2 and 5). 

\subsection{Clinical datasets}
In pathological cases, it is important that our choice of parameters strikes a balance between false negatives (e.g., too little streamlines) and false positives (e.g., generating too many spurious streamlines) in the vicinity of the tumor.

\begin{figure}[htb]
\centering
\includegraphics[width=0.995\textwidth]{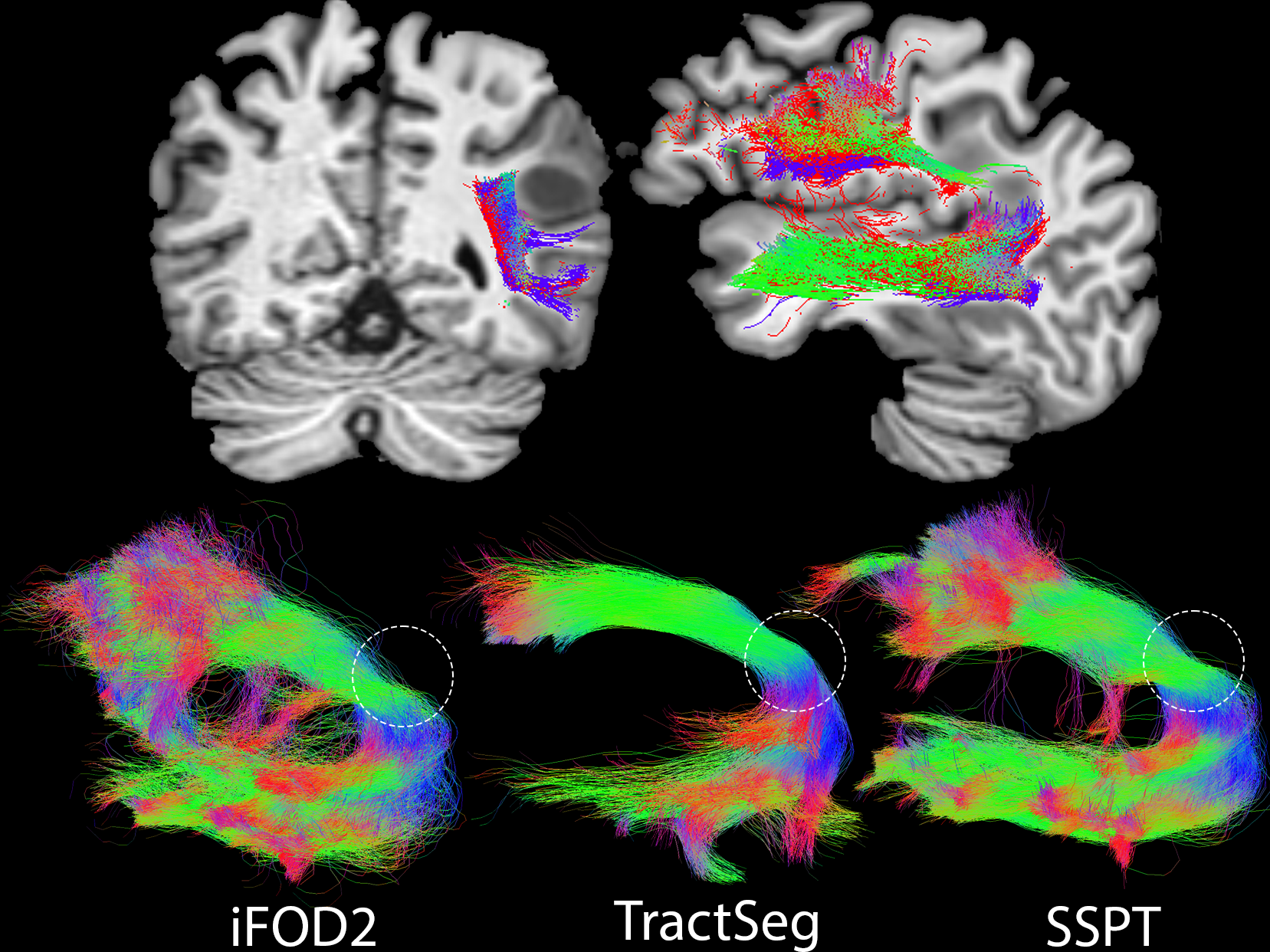}
\caption{All three slices are from the T1 image of patient A with a cross-sectional slab of the tracks of the right AF generated by iFOD2 (red), TractSeg (blue), and SSPT (green). The bottom right shows each result again in a 3D view, with a dashed circle highlighting the location of the tumor.}
\label{fig:AW196}
\end{figure}

\textbf{Patient A:}
Figure~\ref{fig:AW196} shows tracking results of the right AF in patient A. As shown, iFOD2 and SSPT produce a larger volume of bundles than TractSeg. However, SSPT seems to have a higher coverage of endpoint fanning than TractSeg, without the added spurious streamlines of iFOD2. More specifically, in the circle area is a deflection of the bundle, which is less pronounced in TractSeg than in the tracking of iFOD2 and SSPT. 

\begin{figure}[htb]
\centering
\includegraphics[width=0.995\textwidth]{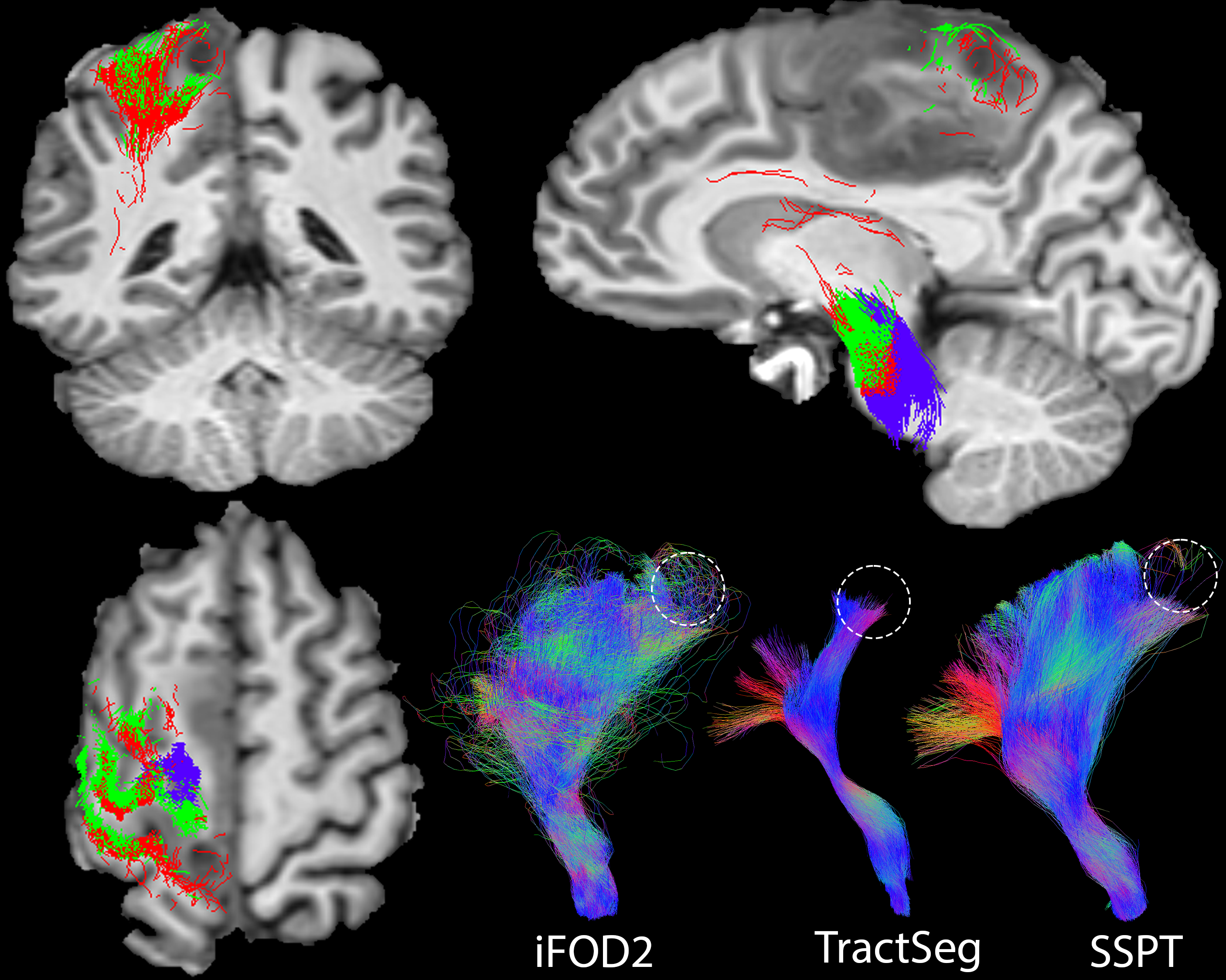}
\caption{The top row shows sagittal and coronal slices of the T1 image of patient B with a cross-sectional slab of the tracks of the left CST generated by iFOD2 (red), TractSeg (blue), and SSPT (green). The bottom row shows the same tracks in a 3D view with a dashed circle highlighting the part of the bundle nearest to the tumor. }
\label{fig:AW199}
\end{figure}

\textbf{Patient B:}
Figure~\ref{fig:AW199} shows the results of tracking the left CST in patient B. One can observe that iFOD2 and SSPT produce a much larger bundle, whereas TractSeg does not produce streamlines that are near the tumor (circled areas). Both iFOD2 and SSPT show curved lines around the tumor, but SSPT produces less spurious streamlines in the tumor area.

\section{Discussion}
\subsection{Experiments}
\subsubsection{Parameter space exploration}
The CST of the HCP dataset shows how the parameter distribution can directly influence what subparts of a bundle are more common in terms of occurrence. 
By looking at the resulting distributions (Figure~\ref{fig:params_all}) of the parameters used to track the entire bundle, one can see that lines with a radius of curvature corresponding to about 3mm generally have a larger chance of resulting in an accepted streamline. 
This same stark contrast is not present in the step size distribution. However, since the larger the step size, the greater the chance of acceptance, the user may be inclined to use as a big of a step size as possible, but this could result in anatomically infeasible streamlines.
This does show that there is not only computational time to be gained by taking larger steps, but for any given seed the resulting streamline is also more likely to be accepted.
The histograms can therefore also be used to selectively adapt the parameter sampling ranges for the respective parameters and increase the likelihood of a seed resulting in an accepted streamline. Doing so speeds up the computation significantly by sampling fewer parameters that have a low probability of finding an accepted streamline. Additionally, Figure~\ref{fig:params_clusters} shows that even though clusters 1 and 5 are spatially very close, their resulting histograms are not. Therefore it is possible to target specific parts of the bundle by analysing the histograms per cluster. 

\subsubsection{Clinical datasets}
Figure~\ref{fig:AW196} showed that iFOD2 seemed to produce more spurious streamlines than both TractSeg and SSPT. 
This is most likely due to the lower FOD threshold used by iFOD2 for the AF. 
TractSeg and SSPT both use the same FOD threshold. 
Therefore it is interesting to investigate the FOD threshold as a variable parameter as well. 
In Figure~\ref{fig:AW199}, TractSeg misses a part of the bundle near the tumor area that the tracks of iFOD2 and SSPT do contain. This is most likely due to the fact that TractSeg is trained on a dataset of healthy subjects and therefore expects normal anatomy.  
This is also a plausible explanation for the less pronounced deflection shown in Figure~\ref{fig:AW196}. 
Using streamline-specific parameters to explore the parameter space would therefore be a viable alternative to deep learning methods trained on healthy patients in pathological cases, since --- even with wide parameter ranges being used --- SSPT behaves the same as iFOD2 with respect to the two tumors presented.

\subsection{Potential applications}
A possible use for parameter space exploration is simplifying the use of tracking algorithms in the clinic since the parameters no longer need to be selected by an expert user, and instead, the domain knowledge of the clinician could suffice.

In the future it would be interesting to step away from spatial clustering and look at similarity between streamlines on a feature level\cite{streamline-similarity-bag-of-features}. On top of that, streamline-specific parametrization could facilitate user-guided real-time tracking by presenting the user with different tracking results and adjusting the parameter ranges on the fly based on their input. Additionally, it might be possible to compare the distributions of the parameters between the left and right side of the brain and see whether the successful parameter combinations possibly contain information regarding abnormalities. The choice of tracking algorithm could also be considered a parameter, and ROI-based parameters could be altered using affine transformations or morphological operators.
Experimentation with other tracking algorithms and parameters will bring further insight into the effectiveness of SSP.

\section{Conclusion}
We have shown that parameter space exploration is a viable method of making more informed parameter selections for tractography. Additionally,  SSPT can find specific parameter combinations that work well for specific parts of certain bundles if an adequate clustering is provided.
This means parameters can be selected with the intent of finding more (or less) of a specific portion of the bundle. A concomitant advantage is that SSPT can be used to speed up tracking by selecting specific parameters.
The downside of this approach is that existing algorithms have to be adapted to take SSP as input, and output the information required for analysis. Additionally, SSPT, as presented, does not consider anatomical plausability when sampling the provided parameter distributions. Even though we have shown two cases in which it behaves the same as iFOD2, this is no guarantee that it will always be correct. \\

\section{Acknowledgements}
This publication is part of the project Bringing Tractography into Daily Neurosurgical Practice with project number KICH1.ST03.21.004 of the research programme Key Enabling Technologies for Minimally Invasive Interventions in Healthcare, which is (partly) financed by the Dutch Research Council (NWO). We would like to thank neurosurgeon Geert-Jan Rutten and scientific programmer Rembrandt Bakker for sharing and preprocessing the clinical dataset used in our experiments at the Elisabeth TweeSteden Hospital (ETZ) in Tilburg, The Netherlands. The authors thank Tom Hendriks for methodological support.
%
%
%
\bibliographystyle{splncs04}
%

\end{document}